\documentclass[aps,pra,nofootinbib,preprint,amsmath,amssymb,floatfix]{revtex4}
\usepackage{color}
\usepackage{graphicx}

\begin{document}

\newcommand{\tc}{\textcolor}
\newcommand{\g}{blue}
\newcommand{\ve}{\varepsilon}
\title{ Viscous coupled fluids in terms of a log-corrected equation of state}         
\author{  I. Brevik$^1$  }      
\affiliation{$^1$Department of Energy and Process Engineering,  Norwegian University of Science and Technology, N-7491 Trondheim, Norway}
\author{K. Myrzakulov$^{2,3}$}
\affiliation{$^2$Eurasian National University, Nur-Sultan 010008, Kazakhstan \\
$^3$Ratbay Myrzakulov Eurasian International Centre for Theoretical Physics, Nur-Sultan 010009, Kazakhstan}
\author{A. V. Timoshkin$^{4,5}$}
\affiliation{$^4$Tomsk State Pedagogical University, Kievskaja Street, 60, 634061 Tomsk, Russia. E-mail: alex.timosh@rambler.ru}
\affiliation{$^5$International Laboratory of Theoretical Cosmology}
\affiliation{Tomsk State University of Control Systems and Radioelectronics, Lenin Avenue, 36, 634050 Tomsk, Russia}
\author{A. Zhadyranova$^{2,3}$ }

\date{\today}          

\begin{abstract}
We consider a class of cosmological fluids that possess properties analogous to those of crystalline solids undergoing isotropic deformations. Our research is based on a modified log-corrected power-law equation of state in the presence of a bulk viscosity. This formalism represents a class of so-called logotropic fluids, and allows explaining an accelerating late-time universe. In order to obtain a more detailed picture of its evolution, we add in our model a coupling of the log-corrected power-law fluid to dark matter, and study various interacting forms between them. We solve the system of equations for a   modified log-power-law fluid coupled to dark matter, and obtain expressions for the log-corrected power-law energy density, and the energy density for dark matter. A comparative analysis is made with the model of a nonviscous log-corrected power-law fluid without interaction with dark matter.

\end{abstract}
\maketitle
Keywords: viscous cosmology; cosmological equation of state;  viscous dark fluid.

Mathematics Subject Classification 2020:  83C55; 83C56.

\bigskip

\section{Introduction}

One of the main tasks in cosmology is to explain the accelerated expansion of the universe \cite{1,2,3}. According to astronomical observations, up to 73\% of the energy density is a component known as dark energy. The remaining 27\% are cold dark matter (CDM), and there is left only 4\% in the form of ordinary baryonic matter concentrated in galaxies and their clusters.  Dark matter is necessary in cosmology to account  for missing mass in the galaxies, and to make the rotation curve flat. In the standard $\Lambda$CDM model, dark matter is pictured as a pressure less fluid. Dark energy does not interact with ordinary matter, and is usually interpreted as a cosmological constant (energy density of a vacuum). The present accelerating expansion of the universe can be explained in terms of an exotic perfect fluid (dark energy) with negative pressure which satisfies a barotropic equation of state  \cite{4,5}. The most general models of dark fluid can be described using an inhomogeneous equation of state \cite{6,7,8}.

As is known, the standard cold dark matter model (CDM) gives good results at the large (cosmological) scale, but has problems at the small (galactic) scale. These problems are related to the assumption that dark matter is usually assumed pressure less. A description of the late-time universe at  small scale can be made with use of the log-corrected power-law equation of state in the Debye approximation \cite{9, 10}. In that model the pressure of the fluid is modeled by an empirical formula giving the   pressure of deformed  isotropic crystalline solids \cite{11}. In order for the universe to change under the cosmological expansion, it is necessary that the pressure of the fluid, described by the equation of state, is negative \cite{12,13,14} (a negative pressure is the same as a positive tensile stress). The negative pressure in the log-corrected power-law model becomes dominant when the volume of the universe exceeds a certain value. This scenario corresponds to the logotropic dark energy model (LDE) \cite{15,16}.  There are regimes in which the log-corrected power-law dark energy is equivalent to a logotropic dark energy.

There exist  many different approaches to modern cosmology. One interesting variant is to make use of machine learning, in order to  restrict the range of central cosmological parameters in a two-component fluid system (this can be useful for elucidating the $H_0$ tension problem too). Such an approach was recently outlined in Ref.~\cite{elizalde21}, where two models were investigated: (i)
  cosmological constant $\Lambda$ + baryonic matter + dark matter (with $\omega_{dm} \neq 0$), and (ii) dark energy (with $\omega_{de} \neq -1$) + baryonic matter + dark matter (with $\omega_{dm} \neq 0).$ Comparison with experimental data gave, as a main result, that $\omega_{dm}
 \neq 0$ was favored in both cases, thus indicating a deviation from the usual cold dark matter model. This phenomenological model is direct and useful.

In the present work we will study the dynamic evolution of a late-time universe, applying a modified log-corrected power-law equation of state in the presence of a bulk viscosity. Our approach is thus relying upon the ansatz (\ref{1}), and is more theoretical in nature than that followed in Ref.~\cite{elizalde21}.
We assume a spatially flat Friedmann-Robertson-Walker space-time, and assume homogeneity and isotropy at a large scale.  To give a  detailed description of the acceleration of the late universe, we adopt a two-component coupled fluid model. The second fluid component is taken to be  pressure less dark matter weakly interacting with the log-corrected power-law fluid. We analyze different types of interaction, and some variants of the bulk viscosity, and obtain solutions of the gravitational equations.

We mention finally some other relevant work. Reference \cite{elizalde20} discusses inhomogeneous single-fluid models for the Universe, also similarly to the above  making use of the machine learning method. Viscous holographic dark energy is discussed in Ref.~\cite{khurshudyan16}.  Interacting holographic dark energy models are discussed in Ref.~\cite{sadri19}. Models of future singularities are classified in Ref.~\cite{elizalde18}, and nonlinear logarithmic interactions are studied in Ref.~\cite{khurshudyan18}.

\section{Modified equation of state for a log-corrected power-law viscous fluid}

Our purpose is to study dark energy in terms of a log-corrected power-law fluid. The equation of state has the form \cite{17, 18}
\begin{equation}
p = A\left( \frac{\rho}{\rho_*}\right)^{-l}\ln \left( \frac{\rho}{\rho_*}\right), \label{1}
\end{equation}
where $\rho_*$ is a reference density, which is identified with the Planck density in \cite{16}: $\rho_p= c^5/\hbar G^2 \approx 5.16\times 10^{99}~$g/m$^3$. In the present notation, $A>0$ represents the  logotropic temperature, while $l= -\frac{1}{6}-\gamma_G$.  For $l=0$ we obtain the equation of state for the logotropic cosmological model \cite{16}. It is interesting to note that the equation of state (\ref{1}) may have a deep relationship to the equation of state considered in Refs.~\cite{10,11}.

Let us rewrite Eq.~(\ref{1}) using the notation of the LDE model. For this purpose we express the volume in terms of the mass density, using the relation $V \sim \rho^{-1}$ \cite{17,18},
\begin{equation}
p(V)= -\beta \left( \frac{V}{V_0}\right)^{ -(1/6)-\gamma_G}\ln \left( \frac{V}{V_0}\right), \label{2}
\end{equation}
where $V_0$ is a volume that distinguishes a barrier between the different signs of the pressure $p$, $\beta$ is the bulk modulus at $V_0$, and $\gamma_G$ is the dimensional Gruneisen parameter. The bulk modulus shows how much the volume changes under the action of external forces. The parameter $\gamma_G$, in the homogeneous and isotropic universe, is a free parameter in the theory. If  $V<V_0$,  the pressure is positive for a positive bulk modulus. If $V>V_0$, the pressure becomes negative for a positive bulk modulus.

If the pressure of the dark fluid satisfies Eq.~(\ref{2}), then in order to ensure the cosmological acceleration the volume must overcome the barrier $V \approx V_0$.  There are three different regimes to be distinguished between \cite{17}:

\noindent 1. The region before passing the $V_0$ barrier, when $V<V_0$. In this region the pressure is positive, and the universe is decelerating. This corresponds to the case of pressure less matter in the $\Lambda$CDM model.

\noindent 2. The region where $V \approx V_0$. In this case there occurs a transition from deceleration to acceleration.

\noindent 3. The region after passing the $V_0$ barrier, when $V>V_0$. Then the pressure is negative (positive tensile stress, as mentioned), and the  fluid starts to accelerate.

Thus, in the log-corrected power-law model the dynamical evolution of the universe is described by one single fluid, which accelerates the universe when its volume passes the barrier $V=V_0$. This allows us to apply this model to the description of the late universe.

We will now study the dynamical  evolution of the universe in some detail, making use of the equation of state (\ref{1}). We assume the universe to be spatially flat, homogeneous and isotropic, and assume there to be a bulk viscosity. We add  the following term to the pressure,
\begin{equation}
\zeta(H,t)= \xi_1(t)(3H)^n, \label{3}
\end{equation}
where $\zeta(H,t)$
is the bulk viscosity, which depends on the Hubble parameter $H$ and the time $t$. From thermodynamical considerations it follows that $\zeta(H,t)>0$. Our augmented equation of state for the log-corrected power-law fluid becomes
\begin{equation}
p = A\left( \frac{\rho}{\rho_*}\right)^{-l}\ln \left( \frac{\rho}{\rho_*}\right) -3H\zeta(H,t). \label{4}
\end{equation}
This is a particular case of a generalized equation of state. To summarize, its basis is the log-corrected power-law fluid equation of state (\ref{1}), in addition to the  term (\ref{3}) that accounts for   bulk viscosity effects in cosmology.

\section{Universe filled with viscous log-corrected coupled fluid}

In this section we will study the late-time universe using the formalism of two viscous coupled fluids in a spatially flat Friedmann-Robertson-Walker space-time. We will  study the dynamics of the accelerating expansion using the log-corrected power-law equation of state (\ref{4}), coupled with dark matter.

Let us clarify the application of the log-corrected power-law model. In this model there is only one single fluid, responsible for the accelerating universe, when the universe volume passes the barrier $V=V_0$. The  model cannot describe the early universe, because the temperature in this period is much higher than the Debye temperature for solids. In the inflationary period, the fluid becomes pressure less as is the case of the LDE model \cite{19}, while at late times the pressure tends to a constant negative value similarly to  the case when the universe is filled with a dark fluid. In our approach where account is taken of viscosity, we have the opportunity to a more accurate picture of the singular behavior of the universe in the vicinity of the Big Rip type I  \cite{20, 21, 22, 23}, or types II, III and IV \cite{24, 25}, the classification of which was first given in \cite{26}.

We consider a universe filled with two interacting components: a log-corrected power-law component having viscosity, and a dark matter component, in a spatially flat Friedmann-Robertson-Walker universe with scale factor $a$. The background equations are \cite{26}
\begin{equation*}
\dot{\rho}+3H(p+\rho)= -Q,
\end{equation*}
\begin{equation}
{\dot{\rho}}_1+3H(p_1+\rho_1)=Q, \label{5}
\end{equation}
and
\begin{equation}
\dot{H}= -\frac{k^2}{2}(p+\rho +p_1+\rho_1). \label{6}
\end{equation}
Here $H(t)= \dot{a}(t)/a(t)$ is the Hubble function, and $k^2=8\pi G$ with $G$ the Newton gravitational constant, $p,\rho$ and $p_1, \rho_1$ are the pressure and the energy density of the coupled fluids, and $Q$ is the interacting term. A dot denotes derivative with respect to the cosmic time $t$. The cosmological constant $\Lambda$ is set equal to zero.

We consider the line element
\begin{equation}
ds^2= -dt^2+a^2(t)\sum_idx_i^2. \label{7}
\end{equation}
The Friedmann for the Hubble parameter is \cite{26}
\begin{equation}
H^2= \frac{k^2}{3}(\rho +\rho_1). \label{8}
\end{equation}
One of the unsolved problems in modern cosmology is the coincidence problem in the standard $\Lambda$CDM model. As the   dark energy density and the  matter energy density are of the same order of magnitude in the present universe, it can be assumed that dark energy and matter somewhat interact with each other. Accurate cosmological observations show that $r = \rho_1/\rho$ is of order unity.

We suppose that the density ration $r$ is constant (the fixed ratio is a consequence of the $\Lambda$CDM model), whereby Eq.~(\ref{7}) is rewritten as
\begin{equation}
\rho= \frac{3H^2}{k^2(1+r)}. \label{9}
\end{equation}
Next, we will investigate cosmological models with various kinds of interaction.

\section{Model with coupling term $Q=3\lambda H(\rho+\rho_1)+\gamma(\dot{\rho}+{\dot{\rho}}_1)$}

We consider the coupling
\begin{equation}
Q=3\lambda H(\rho+\rho_1)+\gamma(\dot{\rho}+{\dot{\rho}}_1), \label{10}
\end{equation}
where the parameters $\lambda$ and $\gamma$ are dimensionless constants. The interaction between the fluid components thus depends on the sign of $\lambda$ and $\gamma$. There is at present no fundamental theory defines the functional form of the coupling; the present assumption is related to that in Ref.~\cite{27}.

We choose the viscosity to be proportional to the Hubble parameter,
\begin{equation}
\zeta(H,t)= 3\tau H, \label{11}
\end{equation}
where $\tau$ is a positive constant.

Using the first member of Eqs.~(\ref{5}) and Eqs.~(\ref{4}), (\ref{8}), (\ref{9}) and (\ref{10}), we obtain the gravitational equation for the viscous log-corrected power-law fluid,
\begin{equation}
\tilde{\gamma}\dot{\rho} +3H\left[ A\left( \frac{\rho}{\rho_*}\right)^{-l}\ln \left( \frac{\rho}{\rho_*}\right) +\theta \rho \right] =0, \label{12}
\end{equation}
where $\theta= 1-(1+r)(\lambda -3\tau k^2)$ and $\tilde{\gamma}= 1+\gamma (1+r)$.

Let us suppose that $\rho > \rho_*/2$ (density of the log-corrected power-law fluid being higher than the Planck density), and let us study the case $l=-1$. Then Eq.~(\ref{12}) simplifies to
\begin{equation}
\tilde{\gamma}\dot{\rho}+\frac{3H}{\rho_*}\left[ A(\rho-\rho_*)+\theta \rho \rho_*\right] =0. \label{13}
\end{equation}
Using Friedmann's equation (\ref{8}) we can rewrite Eq.~(\ref{9}) in the form
\begin{equation}
\tilde{\gamma}\dot{H} +dH^2-b=0, \label{14}
\end{equation}
where $d= \frac{3}{2}(A+\theta \rho_*)$ and $b=\frac{1}{2}A(1+r)k^2\rho_*$.

The solution of Eq.~(\ref{14}) is
\begin{equation}
H(t)= \sqrt{\frac{b}{d}}\, \frac{ \exp{({\tilde{\gamma}}^{-1}\sqrt{bd}\,t}) +C\exp{(-{\tilde{\gamma}}^{-1}\sqrt{bd}\,t})      } {  \exp{({\tilde{\gamma}}^{-1}\sqrt{bd}\,t}) -C\exp{(-{\tilde{\gamma}}^{-1}\sqrt{bd}\,t})   } \label{15}
\end{equation}
where $C$ is an arbitrary constant.

In the particular case when $C=1$ we obtain
\begin{equation}
H(t)=  \sqrt{\frac{b}{d}}\, coth ( {\tilde{\gamma}}^{-1} \sqrt{bd}\, t). \label{16}
\end{equation}
Hence, since $H>0$ the universe is expanding. We see that $H$ diverges for $t \rightarrow 0$, implying a Big Rip type singularity \cite{26}.

The scale factor is given by
\begin{equation}
a(t)= \exp \left[ \int H(t)dt\right] = a_0\sinh ({\tilde{\gamma}}^{-1} \sqrt{bd}\, t)^{\tilde{\gamma}d}, \label{17}
\end{equation}
where $a_0$ an integration constant. We calculate the second derivative of $a(t)$,
\begin{equation}
\ddot{a}(t) =
{\tilde{\gamma}}^{-2}\frac{b}{d}\, \frac{ \cosh^2({\tilde{\gamma}}^{-1} \sqrt{bd}\, t)-d{\tilde{\gamma}}^{-1} }
{  \sinh^2({\tilde{\gamma}}^{-1} \sqrt{bd}\, t)     }a(t).
\end{equation}
Hence, $\ddot{a}(t)=0$ at $t_0= \frac{1}{{\tilde{\gamma}}^{-1}\sqrt{bd}}\ln \left( {\tilde{\gamma}}^{-1/2}\sqrt{d}+\sqrt{{\tilde{\gamma}}^{-1} d-1}           \right)$. In the case when $d<\tilde{\gamma}$ (i.e., if $A<2\tilde{\gamma}/3-\theta \rho_*$), then the second derivative of the scale factor is positive, and the universe experiences an accelerated expansion as is the case at present. On the other hand, in the case when $d>\tilde{\gamma}$ for early times $0<t<t_0$ the first derivative is positive but the second derivative negative, meaning a decelerating expansion, while for $t>t_0$ the universe enters an accelerating era. In this way we are able  to obtain a transition from a dominant matter epoch to a dark energy epoch, in agreement wtth observations. Finally, we note that from Eq.~(\ref{16}),
\begin{equation}
\dot{H}(t)= -\frac{b{\tilde{\gamma}}^{-1}}
{\sinh^2({\tilde{\gamma}}^{-1}   \sqrt{bd}\, t)}, \label{19}
\end{equation}
we deduce, since $\dot{H}<0$, that the universe does not super-accelerate (i.e., the equation-of-state parameter is not phantom).Lastly, we mention that in the case of zero viscosity and no interaction (i.e., for $\tau=0$ and $\lambda=\gamma=0$) we obtain $\theta=\tilde{\gamma}=1$ and so the above analysis is significantly simplified.

WE close this subsection by studying the case of pressure less matter $(p_1=0)$. Then the solution of (\ref{5}) with the coupling term (\ref{10}) and the Hubble function (\ref{15}) takes the form
\begin{equation}
\rho_1(t)= \frac{C_1}{\left[ \sinh^2({\tilde{\gamma}}^{-1}\sqrt{bd}\,t)\right]^{\tilde{\lambda}\tilde{\gamma}/d{\tilde{\gamma}}_*}}, \label{20}
\end{equation}
where ${\tilde{\gamma}}_* = 1-\lambda (1+1/r), \, \tilde{\lambda} = 1-\lambda(1+1/r)$ and $C_1$ is an integration constant. Thus, at late times when $t\rightarrow \infty$ we obtain $\rho_1(t)\rightarrow 0$.

In summary, the model at hand can describe the universe evoluion with transition from a matter epoch to a dark energy epoch, in agreement with observations. We stress that we have not considered a cosmological constant, and thus the above behavior is due to the model dynamics only. This is a significant advantage and one of the main results of the present subsection.

\section{Interacting model with coupling term $Q=3\alpha H\frac{\rho\rho_1}{(\rho+\rho_1)}$ }

Let us suppose that the interaction between the fluid components has the form
\begin{equation}
Q=3\alpha H\frac{\rho\rho_1}{(\rho+\rho_1)}, \label{21}
\end{equation}
where the parameter $\alpha$ is a dimensionless constant. In terms of the energy density ratio $r=\rho_1/\rho$, the interaction term can be written
\begin{equation}
Q= \frac{3\alpha r}{1+r}H\rho. \label{22}
\end{equation}
Let us consider the expression (\ref{3}) for the viscosity in the case $n=1$, and choose the function $\xi_1(t)$ to vary linearly with time,
\begin{equation}
\xi_1(t)= ct+b, \label{23}
\end{equation}
with constant parameters $c,b$.Then the gravitational equation for the log-corrected power-law coupled fluid in the presence of viscosity, under the condition $\rho > \rho_*/2$, will take the form
\begin{equation}
\dot{\rho}+3H\left[ A\left( \frac{\rho}{\rho_*}\right)^{-l}-A\left( \frac{\rho}{\rho_*}\right)^{-l-1}+(\tilde{c}t+\tilde{b})\rho \right] =0, \label{24}
\end{equation}
where $\tilde{c}= -3c(1+r)k^2$ and $\tilde{b}= 1-3b(1+r)k^2+\frac{3\alpha r}{1+r}$.

In the case $l=-1$, using Friedmann's equation (\ref{9}), we can rewrite (\ref{24}) as
\begin{equation}
2\dot{H}+3\left( \tilde{c}t+\tilde{b}+\frac{A}{\rho_*}\right) H^2=0, \label{25}
\end{equation}
whose solution is
\begin{equation}
H(t)= \frac{4\tilde{c}}{3\left( \tilde{c}t+\tilde{b}+\frac{A}{\rho_*}\right)^2+C_2}, \label{26}
\end{equation}
with $C_1$ an integration constant. Without loss  of generality we can focus on the case $C_1=0$. In this case $H>0$ and thus the universe is expanding. Moreover, $H$ diverges at a finite time $t_0= -\frac{1}{\tilde{c}}\left( \tilde{b}+\frac{A}{\rho_*}\right)$, and thus a Big Rip type singularity appears \cite{26}.

The scale factor is given by the expression
\begin{equation}
a(t)= a_0\exp \left[ -\frac{4}{3}\left( \tilde{c}t+\tilde{b}+\frac{A}{\rho_*}\right)^{-1}\right], \label{27}
\end{equation}
where $a_0$ is an integration constant.

The time derivative of the scale factor is
\begin{equation}
\dot{a}(t)= \frac{4/3}{   \left( t+\frac{\tilde{b}}{\tilde{c}} +\frac{A}{\tilde{c}\rho_*}\right)^2} a(t). \label{28}
\end{equation}
The derivative is positive, meaning that the universe expands. The second derivative is
\begin{equation}
\ddot{a}(t)= \left[ 1-\frac{3}{2}\left( \tilde{c}t+ \tilde{b} +\frac{A}{\rho_*}\right) \right]H^2(t)a(t). \label{29}
\end{equation}
Hence, $\ddot{a}(t)=0$ at $t_1= \frac{1}{\tilde{c}}\left( \frac{2}{3\tilde{c}}-\tilde{b}-\frac{A}{\rho_*}\right)$. Thus in the case $\tilde{c}>0$, for values $t<t_1$, it turns out that  $\ddot{a}(t)>0$ and the universe transits to  a late-time accelerated era.

Note that in the case of a non-viscous fluid without interaction with dark matter (parameters $c=b=0$, and $\alpha=0$), in the case $l=-1$, we obtain
\begin{equation}
H(t)= \frac{2/3}{\left( 1+\frac{A}{\rho_*}\right)t +C_3}, \label{30}
\end{equation}
where $C_3$ is an integration constant.

In the particular case when $C_3=0$ the time derivative of $H(t)$ is
\begin{equation}
\dot{H}(t)= -\frac{2}{3\left( 1+\frac{A}{\rho_*}\right)t^2}. \label{31}
\end{equation}
Since $\dot{H}<0$, the universe is decelerating.

The scale factor varies with time as
\begin{equation}
a(t)= a_0  t^{ \frac{2}{3(1+A/\rho_*)}}, \label{32}
\end{equation}
where $a_0$ is an arbitrary constant.

The first derivative of the scale factor is
\begin{equation}
\dot{a}(t)= \frac{2/3}{\left( 1+\frac{A}{\rho_*}\right)t}a(t).
 \label{33}
\end{equation}
The derivative is positive, so that the universe expands.

Thus, without taking into account the viscosity of the log-corrected power-law fluid and its interaction with dark matter, we have obtained a cosmological model that does not describe he accelerating universe expansion at present. This model is therefore less realistic.

The solution of the gravitational equation (\ref{5}) for dark matter, with the coupling term (\ref{22}) and the Hubble function (\ref{26}), has the form
\begin{equation}
\rho_1(t)= {\tilde{\rho}}_0
\exp{ \left[   \frac{A}{\tilde{c}C_4}\left( \frac{\alpha r}{1+r}-1\right)
arctan \left( \frac{\tilde{c}t+\tilde{b}+\frac{A}{\rho_*}}{C_4}\right) \right] },\label{34}
\end{equation}
where $C_4 \neq 0$ is an arbitrary constant and ${\tilde{\rho}}_0 = \rho_1(0)$.

In the limit $t \rightarrow \infty$ for the late universe, we have $\rho_1 \rightarrow {\tilde{\rho}}_0\exp \left[ \frac{2\pi}{\tilde{c}C_4}\left( \frac{\alpha r}{r+1}-1\right)\right]$.

In summary, the recent model can describe the evolution of the  universe in agreement with observations, with a transition from a matter dominated era to a late-time accelerated epoch. It should be observed that this behavior has been obtained without involving a cosmological constant, what is a significant advantage and one of the main results of the present subsection.

\section{Conclusions}

In this article we have considered a dark energy model of the universe, based on the log-corrected power-law modified equation of state (\ref{4}), in the presence of a  bulk viscosity $\zeta$, in a homogeneous and isotropic spatially flat FRW space-time. The log-corrected power-law fluid possesses properties analogous to those of crystalline solids under isotropic deformations, even in cases when the pressure is negative. This formalism allows us to m\emph{\emph{}}odel and explain the accelerating expansion of the late universe in terms of a logotropic dark fluid.  We considered also the coupling with dark matter and obtained analytic expressions for the energy density of both fluids. Based on the expressions for the scale factor $a(t)$ and its first and second derivatives, we identified the regimes of different behavior: either expansion with acceleration (as at the present time), or expansion with deceleration. It was shown that in contrast to the non-viscous log-corrected power-law fluid without interaction with dark matter, our model is  more appropriate for the description of the present universe.

It was previously shown \cite{28} that the interaction between dark energy possessing fluid viscosity, and dark matter, will affect the character of singularities of various types. This circumstance served as a motivation for the present work.

We showed that a cosmological scenario with coupled dark fluids can give rise to a universe that experiences a transition from a matter dominated era to a late-time acceleration era. The agreement with observational data, in particular the Supernovae type Ia (Sn Ia) and the Hubble function, as confirmed in Ref.~\cite{29, 30}, is a consequence of the logarithmic  correction and the viscosity terms.
We thus  investigated various regimes in the accelerated expansion, and  conclude that this description of the late-time universe may lead to interesting results.

In connection with the coupling equations (\ref{5}), we notice that the nature of the coupling term $Q$ is physically unknown. The last of these equations has the form of an energy conservation equation in the presence of a bulk viscosity if $Q>0$, but this picture does not apply to the first equation because viscosity implies heat production. The right hand side would in that case have to be positive.

In the future, it is of interest to study the thermodynamic aspects of the evolution of the  late-time universe \cite{31}, as well as the effect of thermal radiation on the formation of singularity \cite{32}.

\subsection*{Acknowledgments}
The work was  supported by the Ministry of Education and Science of Kazakhstan, Grant  AP08052034.

\end{document}